\newif\ifsingle

\newif\ifFullVersion
\FullVersiontrue 

\ifsingle
\documentclass[11pt,draftclsnofoot, onecolumn]{IEEEtran}		
\else		
\documentclass[10pt,final, twocolumn]{IEEEtran}
\fi

\usepackage{times}
\usepackage{amsmath,dsfont}
\usepackage{amssymb,amsthm}
\usepackage{epsfig,verbatim}
\usepackage{subcaption}
\usepackage{setspace}
\usepackage{color}
\usepackage{cite}
\usepackage{epstopdf}
\usepackage{graphics}
\usepackage{accents}
\usepackage{acronym}
\usepackage{booktabs}
\usepackage{mathtools} 
\usepackage{enumitem}
\usepackage{multirow}
\usepackage{dblfloatfix}

\usepackage[ruled,linesnumbered]{algorithm2e}  

\usepackage
[
        a4paper,
        left=0.625in,
        right=0.625in,
        top=0.75in,
        bottom=1in,
]
{geometry}

\SetKwInput{KwData}{\textbf{Init}} 
\let\oldnl\nl
\newcommand{\nonl}{\renewcommand{\nl}{\let\nl\oldnl}}

\newcommand{\myVec}[1]{{\boldsymbol{#1}}}

\newcommand{\mySet}[1]{\mathcal{#1}}
	
\newcommand{\myY}{{\myVec{Y}}}			 		
\newcommand{\myS}{{\myVec{S}}}			 		
\newcommand{\myState}{\bar{\myS}} 
\newcommand{\myStateR}{\bar{\myVec{s}}}
\newcommand{\HyperParams}{\myVec{\theta}}
\newcommand{\Weights}{\myVec{\varphi}}

\newcommand{\Pdf}[1]{p_{ { #1}} }

\newcommand{\Mem}{L}			 			
\newcommand{\Blklen}{B}			 			
\newcommand{\Blkset}{\mySet{\Blklen}}
\newcommand{\CnstSize}{M}			 			

\ifsingle

\newcommand{\figSpace}{\vspace{-0.2cm}}

\else

\newcommand{\figSpace}{\vspace{-0.2cm}}
\fi 

\acrodef{adc}[ADC]{analog-to-digital convertor}
\acrodef{cs}[CS]{compressed sensing}
\acrodef{dtft}[DTFT]{discrete-time Fourier transform}
\acrodef{dnn}[DNN]{deep neural network} 
\acrodef{csi}[CSI]{channel state information}
\acrodef{bpsk}[BPSK]{binary phase shift keying}
\acrodef{map}[MAP]{maximum a-posteriori probability}
\acrodef{snr}[SNR]{signal-to-noise ratio}
\acrodef{bs}[BS]{base station} 
\acrodef{iot}[IOT]{Interent of Things}
\acrodef{mimo}[MIMO]{multiple-input multiple-output}
\acrodef{mse}[MSE]{mean-squared error}
\acrodef{pdf}[PDF]{probability density function}
\acrodef{rv}[RV]{random variable}
\acrodef{ml}[ML]{machine learning}
\acrodef{fec}[FEC]{forward error correction}
\acrodef{rs}[RS]{Reed-Solomon}
\acrodef{lti}[LTI]{linear time-invariant}
\acrodef{wss}[WSS]{wide-sense stationary}
\acrodef{psd}[PSD]{power spectral density}
\acrodef{ser}[SER]{symbol error rate} 
\acrodef{ber}[BER]{bit error rate} 
\acrodef{gd}[GD]{gradient descent}
\acrodef{sgd}[SGD]{stochastic gradient descent} 
\acrodef{isi}[ISI]{intersymbol interference}  
\acrodef{awgn}[AWGN]{additive zero-mean white real Gaussian noise} 
\acrodef{ut}[UT]{user terminal} 
\acrodef{mmw}[mmWave]{millimeter wave}
\acrodef{noma}[NOMA]{non-orthognal multiple access}
\acrodef{mac}[MAC]{mulitple access channel}
\acrodef{fl}[FL]{Federated learning}
\acrodef{lstm}[LSTM]{long short-term memory}
\acrodef{maml}[MAML]{model-agnostic meta-learning}

\IEEEoverridecommandlockouts 

\title{Meta-ViterbiNet: Online Meta-Learned Viterbi Equalization for Non-Stationary Channels 
}
\author{  
	\IEEEauthorblockN{Tomer Raviv, Sangwoo Park, Nir Shlezinger, Osvaldo Simeone, Yonina C. Eldar, and Joonhyuk Kang\\
	} 
	\thanks{
		This project has received funding from the European Union’s Horizon 2020 research and innovation program under grants No. 646804-ERC-COG-BNYQ, the European Research Council (ERC) under the European Union's Horizon 2020 research and innovation programme (grant agreement No. 725731). It was supported by the Institute of Information $\&$ Communications Technology Planning $\&$ Evaluation (IITP) grant funded by the Korea Government (MSIT) (No.2018-0-00170, Virtual Presence in Moving Objects through 5G) and by the Ministry of Science and ICT (MSIT), South Korea, through the Information
        Technology Research Center (ITRC) Support Program supervised by the Institute of Information and Communications Technology
        Planning and Evaluation (IITP) under Grant IITP-2020-0-01787. Support is also acknowledged from a gift by Huawei Technologies, and from the Israel Science Foundation under grant No. 0100101.
		T. Raviv is   with the School of EE, Tel-Aviv University, Tel-Aviv, Israel (e-mail: tomerraviv95@gmail.com).
		S. Park and O. Simeone are with the Department of Engineering, King’s College London,  U.K. (email: \{sangwoo.park; osvaldo.simeone\}@kcl.ac.uk).
		N. Shlezinger is with the School of ECE, Ben-Gurion University of the Negev, Beer-Sheva, Israel (e-mail: nirshl@bgu.ac.il).
		Y. C. Eldar is with the Faculty of Math and CS, Weizmann Institute of Science, Rehovot, Israel (e-mail: yonina.eldar@weizmann.ac.il). J. Kang is with the School of EE, KAIST, Daejeon, South Korea (e-mail: jhkang@ee.kaist.ac.kr).}

	\vspace{-1.0cm}
	
}
\vspace{-0.75cm}

\begin{document}
	
	\maketitle

	\pagestyle{empty}
	\thispagestyle{empty}
	\begin{abstract} 
	Deep neural networks (DNNs) based digital receivers can potentially operate in complex environments. However, the dynamic nature of communication channels implies that in some scenarios, DNN-based receivers should be periodically retrained in order to track temporal variations in the channel conditions. To this aim, frequent transmissions of lengthy pilot sequences are generally required, at the cost of substantial overhead. In this work we propose a DNN-aided symbol detector, Meta-ViterbiNet, that tracks channel variations with reduced overhead by integrating three complementary techniques: $1)$ We leverage domain knowledge to implement a model-based/data-driven equalizer, ViterbiNet, that operates with a relatively small number of trainable parameters; $2)$ We tailor a meta-learning procedure to the symbol detection problem, optimizing the hyperparameters of the learning algorithm to facilitate rapid online adaptation; and $3)$ We adopt a decision-directed approach based on coded communications to enable online training with short-length pilot blocks. Numerical results demonstrate that Meta-ViterbiNet operates accurately in rapidly-varying channels, outperforming the previous best approach, based on ViterbiNet or conventional recurrent neural networks without meta-learning, by a margin of up to 0.6dB in bit error rate in various challenging scenarios.
	\end{abstract}
	\vspace{-0.4cm}
	\section{Introduction}
\vspace{-0.1cm} 
	
Deep learning systems have demonstrated unprecedented success in various applications, ranging from computer vision to natural language processing, and recently also digital communications and receiver design \cite{gunduz2019machine,simeone2018very,balatsoukas2019deep, oshea2017introduction}. While traditional receiver algorithms are model-based, relying on mathematical modeling of the signal transmission, propagation, and reception, \acp{dnn} are model-agnostic, and are trained from data. \ac{dnn}-aided receivers can operate efficiently in scenarios where the  channel model is unknown, highly complex, or difficult to optimize for~\cite{farsad2018neural}.

Despite its potential in implementing digital receivers \cite{shlezinger2020inference,farsad2020data}, deep learning solutions are subject to several challenges that limit their applicability in important communication scenarios. A fundamental difference between digital communications and traditional deep learning applications stems from the dynamic nature of communication systems, and particularly of wireless channels.  \acp{dnn} consist of highly-parameterized models that can represent a broad range of mappings. As such, massive data	sets are typically required to learn a desirable mapping.  The dynamic nature of communication channels implies that the statistical model can change considerably over time, and thus a \ac{dnn} trained for a given channel may no longer perform well on a future channel. \ac{dnn}-aided receivers are thus likely to require frequent retraining, at the cost of degraded spectral efficiency due to pilot transmissions. 

Various strategies have been proposed in the literature to facilitate the application of \acp{dnn} to receiver design in dynamic channel conditions. The first type avoids retraining, attempting instead to learn a single mapping that is applicable to a broad range of channel conditions. This class of methods includes the straightforward approach of training a \ac{dnn} using data corresponding to a broad set of expected channel conditions, which is commonly referred to as {\em joint learning}\cite{oshea2017introduction, xia2020note}.  Alternatively, one can train in advance a different network for each expected statistical model, and combine them as a deep ensemble \cite{raviv2020data}. However, these strategies typically require large training data, and  deviating from the training setup can greatly harm performance \cite{simeone2020learning}.

The alternative strategy is to periodically retrain the network. To provide data for retraining, one must either transmit frequent pilots, or, alternatively, use decoded data for training. Such self-supervised training can be implemented by either using successfully decoded \ac{fec} codewords, as  in \cite{shlezinger2019viterbinet, teng2020syndrome}, or  by providing a measure of confidence per each symbol and selecting those with the highest confidence for retraining, as proposed in \cite{sun2020generative}. Nonetheless,  the volumes of data one can obtain in real-time, either from pilots or from decoded transmissions, are limited and are not at the scale of typical data volumes used for training \acp{dnn}. Retrained \ac{dnn}-aided receivers should thus utilize compact \ac{dnn} architectures. This can be achieved without compromising accuracy by using hybrid model-based/data-driven receivers, that incorporate domain knowledge. Following this principle, data-driven implementations of the Viterbi scheme \cite{viterbi1967error},  BCJR method \cite{bahl1974optimal}, and  iterative soft interference cancellation  \cite{choi2000iterative} were proposed in \cite{shlezinger2019viterbinet,shlezinger2020data, shlezinger2019deepSIC}, respectively.

\begin{figure*}[!ht]
    \centering
    \includegraphics[width=0.67\textwidth,height=0.1\textheight]{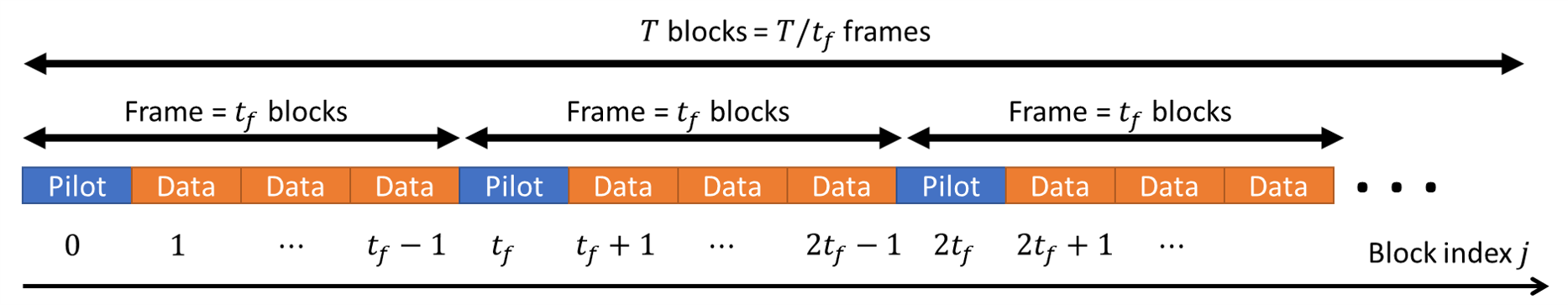}
    \caption{Transmission model. The channel is constant within each block and  changes across blocks, i.e., $ t_f$ times within a frame. }
    \figSpace
    \label{fig:transmission}
\end{figure*}

The ability to retrain quickly is highly dependent on the selection of a suitable initialization of the iterative training algorithm. While the common strategy is to use random weights, the work \cite{shlezinger2019viterbinet} used the previous learned weights as an initial point for  retraining. 
An alternative approach is to optimize the initial point via meta-learning \cite{park2020meta,simeone2020learning,jiang2019mind,park2020learning}. Following this approach, one not only retrains, but also optimizes the hyperparameters that dictate the retraining process. In particular, it was shown in \cite{park2020learning} that by optimizing the initial weights used in the training algorithm, rather than using random weights or the current ones, the receiver can quickly adapt to varying channel-conditions. 


In this work we propose Meta-ViterbiNet, which is a hybrid model-based/data-driven symbol detection mechanism for finite-memory channels, that is capable of tracking time-varying conditions quickly and with minimal overhead. Meta-ViterbiNet enables \ac{dnn}-aided equalization with rapid retraining by combining dedicated designs of the system architecture, training algorithm, and data used for training: 
\begin{itemize}
    \item \textbf{Architecture -} Meta-ViterbiNet employs the ViterbiNet architecture proposed in \cite{shlezinger2019viterbinet}, leveraging domain knowledge about optimal detectors for  finite-memory channels in the presence of \ac{csi} to reduce the number of trainable parameters. 
    \item \textbf{Training algorithm -} We tailor the \ac{maml} method \cite{finn2017model} to incorporate temporal evolution over a sequence of symbols. The goal is to optimize the initialization of the training algorithm, such that training on the last decoded data block minimizes the error on the next data block.
    \item \textbf{Data -} Apart from the pilots, the data used for training is acquired from the local \ac{fec} decoder as in \cite{shlezinger2019viterbinet,teng2020syndrome}, enabling the use of self-generated labels that extend the availability of supervised data beyond the pilot blocks. 
\end{itemize}

 The rest of this paper is organized as follows:  Section~\ref{sec:Model} details the system model. 
 Section~\ref{sec:MetaViterbiNet} presents Meta-ViterbiNet. Experimental results are presented in Section~\ref{sec:Simulation}. Finally, Section~\ref{sec:Conclusions} provides concluding remarks. 	 

Throughout the paper, we use boldface letters for vectors, e.g., ${\myVec{x}}$;
the $i$th element of ${\myVec{x}}$ is written as $({\myVec{x}})_i$. We use upper-case letters for \acp{rv}, and lower-case letters for deterministic quantities. Calligraphic letters, such as $\mySet{X}$, are used for sets, 
and $\mySet{R}$ is the set of real numbers.

	\vspace{-0.2cm}
	\section{System Model}
\label{sec:Model}
\vspace{-0.1cm}

Here, we describe the system model for which Meta-ViterbiNet is designed. We first detail the time-varying channel model in Subsection~\ref{subsec:Channel}, after which we discuss the transmission model and formulate the problem in Subsection~\ref{subsec:Problem}. 	

\vspace{-0.2cm}
\subsection{Channel Model}
\label{subsec:Channel}
\vspace{-0.1cm} 
We consider communications over causal finite-memory blockwise-stationary  channels. Accordingly, the channel output depends on the last $\Mem >0$ transmitted symbols, where $\Mem$ is the memory length. The channel is constant within a block of $\Blklen$ channel uses, which corresponds to the coherence duration of the channel. Let $S_{i,j}\in\mySet{S}$, with $|\mySet{S}|=\CnstSize$, be the symbol transmitted from constellation $\mySet{S}$ at the $i$th time instance $i \in \{1,2,\ldots, \Blklen\}:= \Blkset$ of the $j$th block. The corresponding channel output, denoted $\myY_{i,j}$, is given by a stochastic function of the last $L$ transmitted symbols $\myState_{i,j} := [S_{i-\Mem+1,j},\ldots, S_{i,j}]^T$.  Specifically, by defining the $j$th transmitted block as $\myS^{\Blklen}_j := \{S_{i,j}\}_{i\in \Blkset}$ and its corresponding observations as $\myY^{\Blklen}_j:= \{\myY_{i,j}\}_{i\in \Blkset}$,  the conditional distribution of the channel output given its input satisfies   
\begin{equation}
	\label{eqn:ChModel1}
	\Pdf{\myY^{ \Blklen}_j | \myS^{ \Blklen}_j}\left(\myVec{y}^{ \Blklen}_j | \myVec{s}^{ \Blklen}_j \right)  = 
	\prod\limits_{i\!=\!1}^{\Blklen}\Pdf{\myY_{i,j} | \myState_{i,j}}\left(\myVec{y}_{i,j} |\myStateR_{i,j}\right).  
\end{equation}
In \eqref{eqn:ChModel1}, the lower-case notations $\myVec{y}_{i,j}$ and $\myStateR_{i,j}$ represent the realizations of the \acp{rv} $\myY_{i,j}$ and $\myState_{i,j}$, respectively. We set $S_{i,j} \equiv 0$ for $i<0$, i.e., we assume a guard interval at least $\Mem$ time instances between blocks. Each symbol $S_{i,j}$ is uniformly distributed over the set $\mySet{S}$ of $\CnstSize$ constellation points.
	
\vspace{-0.2cm}
\subsection{Problem Formulation}
\label{subsec:Problem}
\vspace{-0.1cm}  
 We consider the transmission scenario illustrated in Fig.~\ref{fig:transmission}, where a total of $T$ blocks, indexed  $j\in\{0,\ldots,T-1\}$, are transmitted sequentially. Each consecutive $t_f$ blocks constitute a \textit{frame}; e.g., the first frame is comprised of blocks $j\in\{0,\ldots,t_f-1\}$. The first block  of each frame is a known \textit{pilot}, while the remaining $t_f-1$ blocks contain coded data. We denote the set of pilot blocks indices as $\mathcal{J}_p = \{n \cdot t_f| n\in \mathbb{N} \}$. Each coded data block $\myVec{s}^{ \Blklen}_j$ of $B$ symbols conveys a $k$ bit random message $\boldsymbol{m}_j\in\{0,1\}^k$, encoded 
  %
 using both \ac{fec} coding and error detection codes.
 Error detection codes, such as cyclic redundancy check, allow the receiver to determine if decoding of the message $\boldsymbol{m}_j$ is successful or erroneous. 
 
Our goal is to design a symbol detection mechanism for recovering the data symbols. A symbol detector can be written as a mapping $\hat{\myVec{s}}_j^\Blklen:\mySet{Y}^\Blklen \mapsto \mySet{S}^\Blklen$, and the design objective is the symbol error rate on the data blocks, i.e.,
\begin{equation}
\label{eqn:ErrorRate}
	\frac{1}{\Blklen}\sum_{i=1}^{\Blklen} \Pr\left( \hat{s}_{i,j}(\myY^\Blklen_j) \neq  S_{i,j} \right), \qquad j \notin \mySet{J}_p. 
\end{equation}
\vspace{-0.7cm}
	\vspace{-0.2cm}
	\section{Meta-ViterbiNet}
	\label{sec:MetaViterbiNet}
	\vspace{-0.1cm}
	In this section we present Meta-ViterbiNet, which is a \ac{dnn}-aided receiver architecture for time-varying finite-memory channels. We describe the different components of Meta-ViterbiNet in Subsection~\ref{subsec:high-level}. Then, we elaborate on its main components, which are the ViterbiNet architecture, codeword-level online training, and	the meta-learning process, in Subsections~\ref{subsec:ViterbiNet}, \ref{subsec:Online}, and \ref{subsec:Meta}, respectively. 

\vspace{-0.2cm}
\subsection{High-Level Description} 
\label{subsec:high-level}
\vspace{-0.1cm}
Meta-ViterbiNet operates without explicit knowledge of the channel input-output relationship \eqref{eqn:ChModel1}, apart from its memory $L$ and its coherence time duration. The detector for the $j$th block is parameterized by the weight vector $\Weights_j$. In order to enable an adaptation mechanism, the receiver maintains at each block index $j$ a vector of hyperparameters $\HyperParams_j$, as well as a labelled data buffer $\mathcal{D}_j$. This buffer contains pairs of previously received blocks $\myVec{y}_j^\Blklen$ along with their corresponding transmitted signal $\myVec{s}_j^\Blklen$, or an estimated version thereof. The buffer $\mathcal{D}_j$ contains $D$ such pairs, and is managed in a first-in-first-out mode. Following the  \ac{maml} approach \cite{finn2017model}, the hyperparameter vector $\HyperParams_j$ determines the initialization used to update the detector's parameters $\Weights_{j}$ for block $j$ via \ac{sgd} based on recent data.

\begin{figure}
    \centering
    \includegraphics[width=\columnwidth]{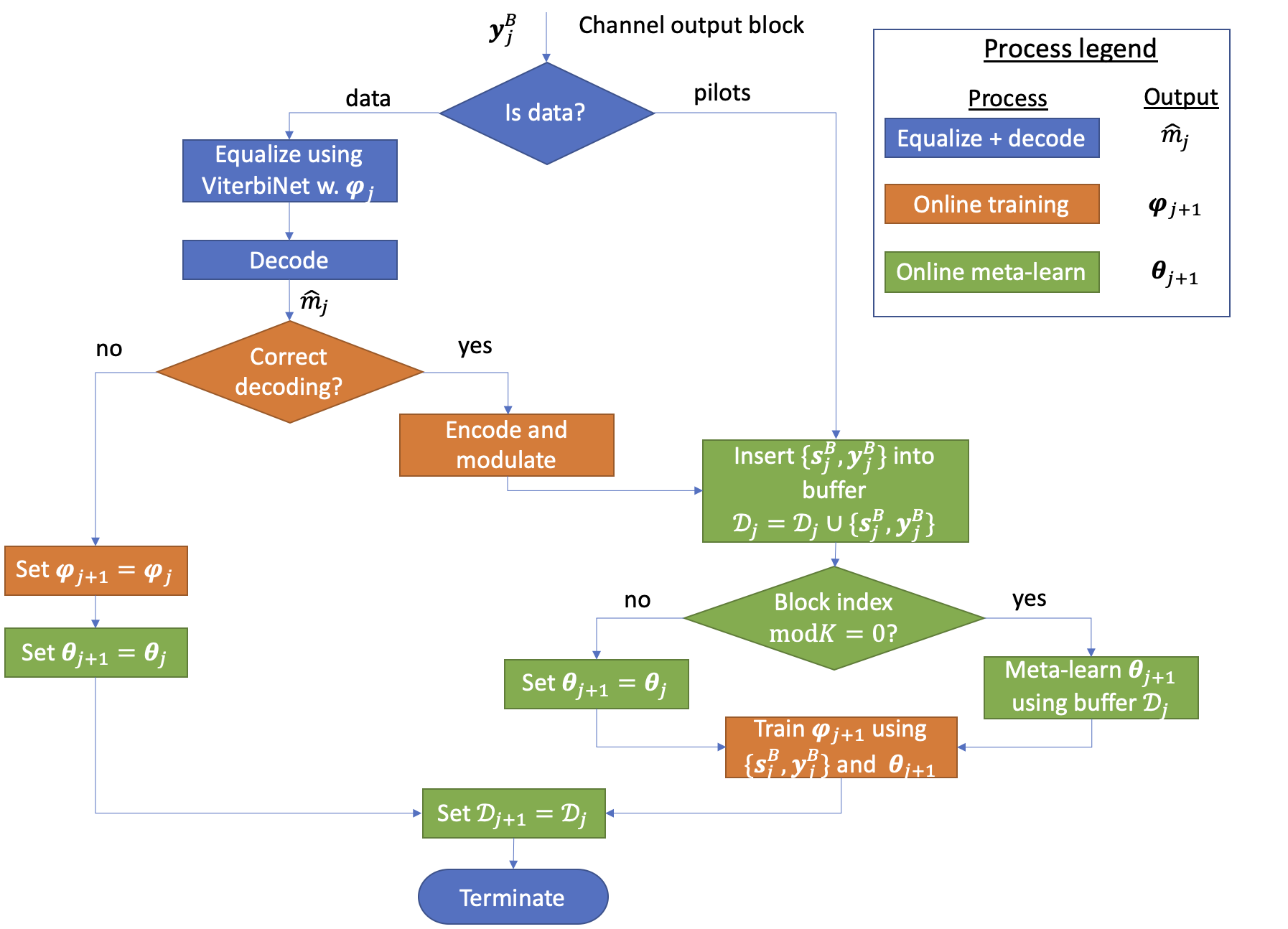}
    \caption{Illustration of the operation of Meta-ViterbiNet.}
    \label{fig:FlowChart1}
\end{figure}
	
As illustrated in Fig. \ref{fig:FlowChart1}, upon the reception of a block of channel outputs $\myVec{y}_j^B$, Meta-ViterbiNet operates in three stages:
\begin{enumerate}
    \item \emph{Detection}: Each incoming data block $\myVec{y}_j^B$ is first equalized by using the ViterbiNet equalizer parametrized by the current vector $\Weights_{j}$, as detailed in Subsection~\ref{subsec:ViterbiNet}. Then, it is decoded by using an arbitrary hard-input \ac{fec} decoder to produce the decoded message $\hat{\myVec{m}}_j$. When decoding is correct, as determined by error detection, the message $\hat{\myVec{m}}_j$ is re-encoded and modulated, producing an estimated transmitted vector $\myVec{s}_j^B$. This block is inserted along with its observations $\myVec{y}_j^B$ into buffer $\mathcal{D}_{j}$. A pilot block $(\myVec{s}_j^B, \myVec{y}_j^B)$ is directly inserted into  $\mathcal{D}_{j}$ upon reception.
    \item \emph{Online training}: In each data block $j$, if decoding is successful, the weights of ViterbiNet $\Weights_{j+1}$ are updated by using the hyperparameters $\HyperParams_{j+1}$ and the newly decoded block $(\myVec{s}_{j}^B, \myVec{y}_{j}^B)$, as detailed in Subsection~\ref{subsec:Online}. Otherwise, no update is carried out. A similar update takes place for pilot block $j$ with pilot block $(\myVec{s}_j^B, \myVec{y}_j^B)$.
    \item \emph{Online meta-learning}: Periodically, i.e., once every $K$ blocks, the buffer $\mathcal{D}_j$ is used to update $\HyperParams_{j+1}$ via online meta-learning, as detailed in Subsection~\ref{subsec:Meta}.
\end{enumerate} 

If $\Weights_j$ and/or $\HyperParams_j$ are not updated in a given block index $j$, they are preserved for the next block by setting  $\Weights_{j+1} = \Weights_j$ and/or $\HyperParams_{j+1}=\HyperParams_j$.
The online adaptation framework is detailed in the sequel, and is summarized in Algorithm~\ref{alg:online-meta-learning}.

\begin{algorithm} 
\DontPrintSemicolon
\caption{Online Adaptation on Incoming Block $j$}
\label{alg:online-meta-learning}
\KwIn{Step sizes $\eta, \kappa$;  threshold $\epsilon$; buffer $\mathcal{D}_{j}$; hyperparameter $\HyperParams_{j}$}
\KwOut{Hyperparameter $\HyperParams_{j+1}$; weights $\Weights_{j+1}$; buffer $\mathcal{D}_{j+1}$}
\vspace{0.15cm}
\hrule
\vspace{0.15cm}
Receive ${\myVec{y}}_j^\Blklen$ 	\tcp*{received channel output}
\uIf{$j\in\mathcal{J}_p$}{
$\mathcal{D}_{j} \leftarrow \mathcal{D}_j \bigcup\{\myVec{s}^\Blklen_j, {\myVec{y}}_j^\Blklen\}$ \tcp*{known pilots} 
  }
  \Else{
  Equalize and decode ${\myVec{y}}_j^\Blklen$ into  $\hat{\myVec{m}}_j$ \tcp*{data}
  \uIf{Decoding is correct}{
  Modulate  $\hat{\myVec{m}}_j\mapsto\myVec{s}^\Blklen_j$ \; 
$\mathcal{D}_{j} \leftarrow \mathcal{D}_j \bigcup\{\myVec{s}^\Blklen_j, {\myVec{y}}_j^\Blklen\}$ 
  }
  }
\nonl\texttt{Online meta-learning (every $K$ blocks)}\;
Set $\HyperParams_{j+1}^{(0)} = \HyperParams_j$\;
\For{$i=1,2,\ldots$}{
\label{stp:Random} Randomly select block $\{\myVec{s}^\Blklen_{\hat{j}+1}, {\myVec{y}}_{\hat{j}+1}^\Blklen\}\in \mathcal{D}_{j}$\; 
      {\uIf{$\{\myVec{s}^\Blklen_{\hat{j}}, {\myVec{y}}_{\hat{j}}^\Blklen\}\notin \mathcal{D}_{j}$}{
  go back to line \ref{stp:Random} \tcp*{invalid data for meta-learning}
  }
  }
Locally update ViterbiNet equalizer for block $\hat{j}+1$ with selected block $\{\myVec{s}^\Blklen_{\hat{j}}, {\myVec{y}}_{\hat{j}}^\Blklen\}$ via \eqref{eq:local_update} as 
  \begin{equation*}
\hat{\Weights}_{\hat{j}+1} = \HyperParams_{j+1}^{(i)}-\eta\nabla_{\HyperParams_{j+1}^{(i)}} \mySet{L}_{\hat{j}}(\HyperParams_{j+1}^{(i)}). \end{equation*} \;
Evaluate loss at block $\hat{j}+1$
, $\mySet{L}_{\hat{j}+1}(\hat{\Weights}_{\hat{j}+1})$\;
Update hyperparameter $\HyperParams_{j+1}$ \tcp*{meta-update}
  \begin{equation*}
\HyperParams_{j+1}^{(i+1)} = \HyperParams_{j+1}^{(i)} - \kappa \nabla_{\HyperParams_{j+1}^{(i)}}\mySet{L}_{\hat{j}+1}(\hat{\Weights}_{\hat{j}+1}). 
\end{equation*}
}
Set hyperparameter $\HyperParams_{j+1} =\HyperParams_{j+1}^{(i+1)}$ \;
\nonl\texttt{Online learning (on each block)} \;
{\uIf{($j\in\mathcal{J}_p$) or (Decoding is correct)}{
 Train $\Weights_{j+1}$ with $\{{\myVec{s}}^\Blklen_j, {\myVec{y}}_j^\Blklen\}$ and initialization $\HyperParams_{j+1}$\;  via \eqref{eq:meta_1} \; 
  }
  \Else{ 
      $\Weights_{j+1} \leftarrow \Weights_j$ \tcp*{no update}
  }
  $\mathcal{D}_{j+1}\leftarrow \mathcal{D}_{j}$ \tcp*{keep buffer}
  }
\end{algorithm}

\vspace{-0.2cm}
\subsection{ViterbiNet Symbol Detection}
\label{subsec:ViterbiNet}
\vspace{-0.2cm} 
The ViterbiNet equalizer, proposed in \cite{shlezinger2019viterbinet}, is a data-driven implementation of the Viterbi detector for finite-memory channels of the form \eqref{eqn:ChModel1} \cite{viterbi1967error}. ViterbiNet does not require prior knowledge of the channel conditional distributions $ \Pdf{\myVec{Y}^{ \Blklen}_j | \myVec{S}^{ \Blklen}_j}$. 

For a given data block $j$, the Viterbi equalizer solves the maximum likelihood sequence detection problem
\begin{align}
\hat{\myVec{s}}^{ \Blklen}_j\left( \myVec{y}_j^{ \Blklen}\right)   
&= \mathop{\arg \min}_{\myVec{s}^{ \Blklen} \in \mySet{S}^\Blklen }\left\{-\sum\limits_{i=1}^{\Blklen } \log \Pdf{\myVec{Y}_{i,j} | \myState_{i,j}}\left( \myVec{y}_{i,j}  |  \myStateR_{i,j}\right)\right\}.
\label{eqn:ML3} 
\end{align}	
In particular,  \eqref{eqn:ML3} is solved recursively via dynamic programming, by  iteratively updating a {\em path cost} $c_i(\myStateR)$ for each state $\myStateR\in \mySet{S}^{\Mem}$ for $i=1,2,\ldots,B$. ViterbiNet implements Viterbi detection in a data-driven fashion by training a \ac{dnn} to provide a parametric estimate of the likelihood function $\Pdf{\myVec{Y}_{i,j} | \myState_{i,j}}\left( \myVec{y}  |  \myStateR\right)$, which is denoted as $\hat{P}_{\Weights}\left( \myVec{y}  |  \myStateR\right)$, where $\Weights$ are the model parameters. See \cite{shlezinger2019viterbinet} for more details.

\vspace{-0.4cm}
\subsection{Self-Supervised Online Training}
\label{subsec:Online}
\vspace{-0.1cm} 	
During data block $j\notin\mySet{J}_p$, the channel decoder takes as input the estimated block $\hat{\myVec{s}}^{ \Blklen}_j$ from the ViterbiNet equalizer, and outputs a decoded message $\hat{\myVec{m}}_j$ along with an indication on the correctness of its decoded message. When decoding is correct, the decoded message $\hat{\myVec{m}}_j$ is encoded and modulated into the estimated transmitted symbols  $\myVec{s}^\Blklen_j$.

At each data block $j$, given the current initialization hyperparameter vector $\HyperParams_{j+1}$ (discussed in the next subsection) and the last successfully decoded block $\{\myVec{s}^\Blklen_{\hat{j}}, \myVec{y}^\Blklen_{\hat{j}}\}$, the algorithm updates the model parameters vector $\Weights_{j+1}$ by minimizing the empirical cross entropy loss:
\begin{align}
\label{eq:meta_1}
  \mathop{\arg \min}_{\Weights} \left\{ \mySet{L}_{\hat{j}}(\Weights) = -\sum_{i=1}^{\Blklen}\log \hat{P}_{\Weights}\left(\myVec{y}_{i,\hat{j}}|\myStateR_{i,\hat{j}} \right)\right\}.
\end{align}
 The optimization problem in \eqref{eq:meta_1} is approximately solved via \ac{gd}, i.e., through iterations of the form
 \begin{align}
\label{eq:local_update}
\Weights_{j+1} = \HyperParams_{j+1} - \eta \nabla_{\HyperParams_{j+1}}\mySet{L}_{\hat{j}}(\HyperParams_{j+1}),
 \end{align}
 where $\eta > 0$ is the learning rate. We note that the index $\hat{j}$ of the last decoded block may be smaller than $j$. While \eqref{eq:local_update} describes a single \ac{gd} iteration, multiple iterations are similarly accommodated. Note also that stochastic computation of the gradient in \eqref{eq:local_update} can be achieved via random sampling among available $B$ blocks to implement stochastic GD (\ac{sgd}).
 

\vspace{-0.4cm}
\subsection{Meta-Learning the Initial Weights}
\label{subsec:Meta}
\vspace{-0.1cm} 	
The hyperparameter {$\HyperParams_{j+1}$} should be optimized so as to enable fast and efficient adaptation of the model parameter $\Weights_{j+1}$ based on the last successfully decoded block $\{\myVec{s}^\Blklen_{\hat{j}}, {\myVec{y}}_{\hat{j}}^\Blklen\}$ using \eqref{eq:local_update}. Adopting MAML \cite{finn2017model}, we leverage the data in the buffer $\mathcal{D}_j$ 
\begin{figure*}
    \centering
    \begin{subfigure}[b]{0.32\textwidth}
    \includegraphics[width=\textwidth,height=0.15\textheight]{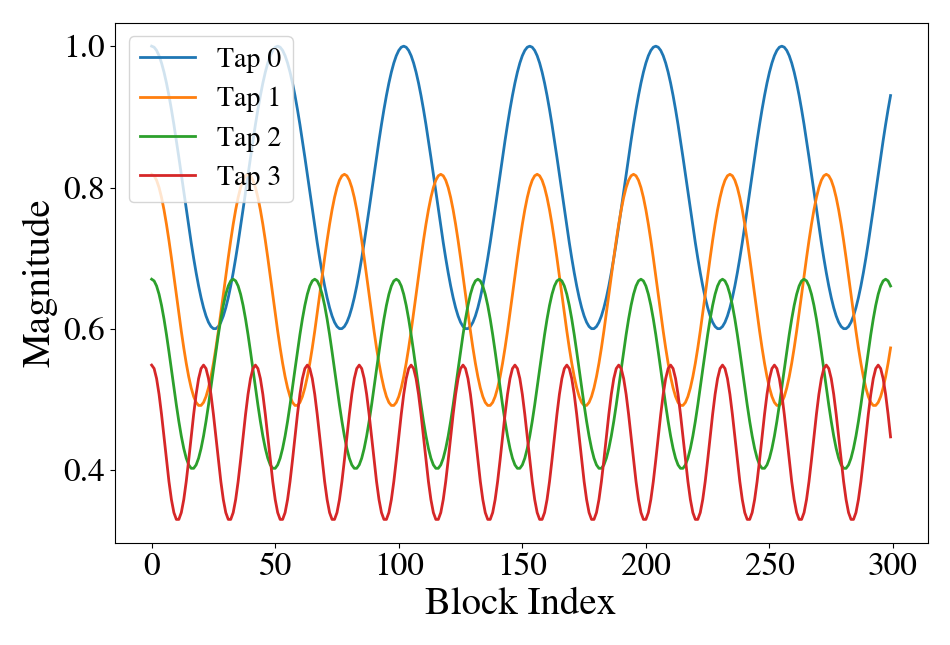}
    \caption{Synthetic preliminary training channel.}
    \label{fig:synthetic_train_channel}
    \end{subfigure} 
    \begin{subfigure}[b]{0.32\textwidth}
    \includegraphics[width=\textwidth,height=0.15\textheight]{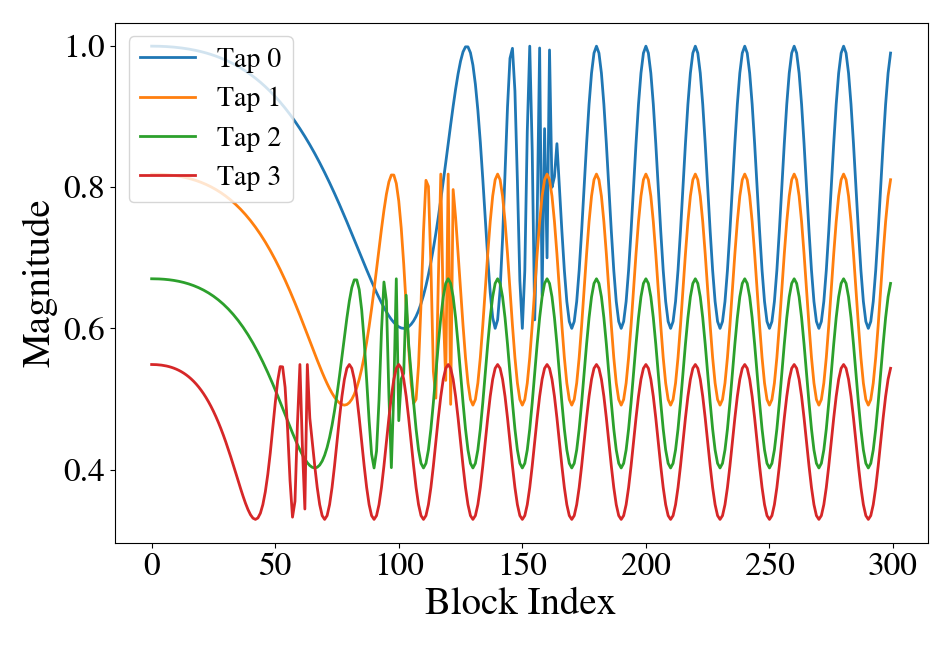}
    \caption{Synthetic test channel.}
    \label{fig:synthetic_test_channel}
    \end{subfigure}
    \begin{subfigure}[b]{0.32\textwidth}
    \includegraphics[width=\textwidth,height=0.15\textheight]{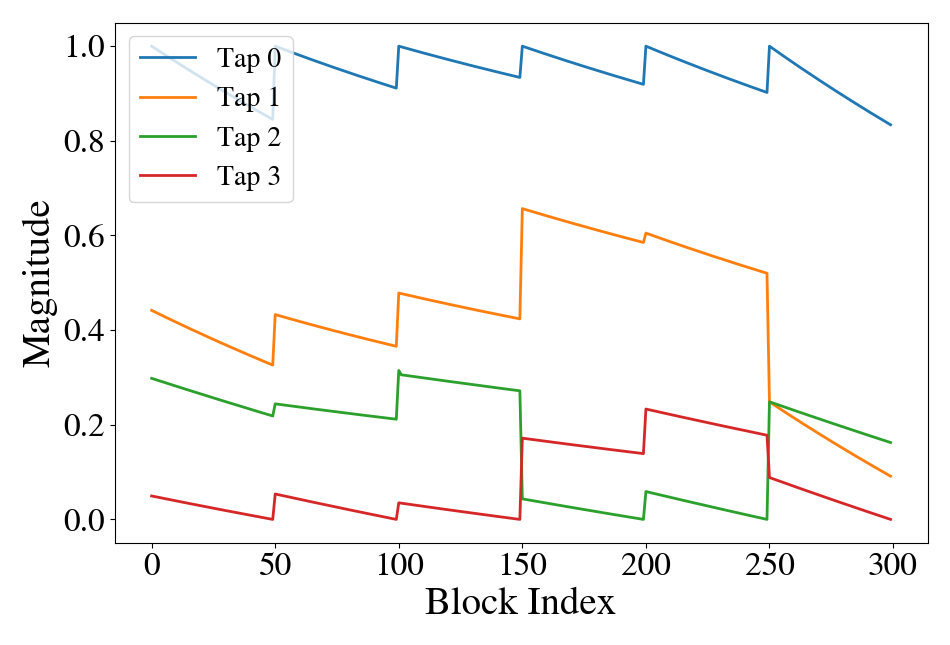}
    \caption{COST 2100 channel.}
    \label{fig:COSTChannel} 
    \end{subfigure}
    \caption{Examples of time-varying channels: channel coefficients versus block index.}
    \figSpace
    \label{fig:channels} 
    \vspace{-0.18cm}
\end{figure*}
%
by considering the problem
\begin{align}
    \label{eq:meta_2}
     \!\HyperParams_{j\!+1} \!=\! \mathop{\arg \min}_{\HyperParams}\!\!\!\! \sum_{\{\myVec{s}^\Blklen_{\hat{j}+1}, {\myVec{y}}_{\hat{j}\!+1}^\Blklen\} \in \mathcal{D}_j}\!\!\! \!\mySet{L}_{\hat{j}\!+1}(\Weights_{\hat{j}+1}\!= \!\HyperParams \!-\! \kappa \nabla_{\HyperParams}\mySet{L}_{\hat{j}}(\HyperParams)),
\end{align}
where $\kappa >0$ is the meta-learning rate. The parameters $\Weights_{\hat{j}+1}$ in \eqref{eq:meta_2} follow same update rule in \eqref{eq:local_update} by using the last available block $\{\myVec{s}_{\hat{j}}^B, \myVec{y}_{\hat{j}}^B\}$ in the buffer  prior to index $\hat{j}+1$. Furthermore, in line with \eqref{eq:meta_1}, the loss $\mySet{L}_{\hat{j}+1}(\Weights_{\hat{j}+1})$ is computed based on data from the following available block $\{\myVec{s}_{\hat{j}+1}^B, \myVec{y}_{\hat{j}+1}^B\}$. When the buffer $\mathcal{D}_j$ contains a sufficiently diverse set of past channel realizations, the hyperparameter obtained via \eqref{eq:meta_2} should facilitate fast training for future channels via \eqref{eq:local_update} \cite{park2020end}. 

{\bf Discussion:} Meta-ViterbiNet is designed to exploit partial domain knowledge regarding both the channel and the transmission protocol in order to enable quick online training with minimal overhead. In particular, the finite memory of the channel allows the use of compact \acp{dnn} without compromising detection accuracy via the ViterbiNet architecture. 
Furthermore, the initial weights of the learning algorithm are periodically updated via online meta-learning to allow fast re-training. By meta-learning over subsequent pairs, in a manner that follows the online retraining procedure,  the detector learns initial weights from which it can rapidly train based on a buffer of past data. This further reduces the amount of data needed to adapt the detector as compared  to which used the the last parameter vector $\Weights_i$ to initialize the update of $\Weights_{i+1}$.


The current formulation of the online adaptation mechanism accounts only for pilot and data blocks. In practice, communication protocols induce additional structures not considered in our design, such as the presence of headers and management frames, which can also be utilized to generate data for retraining. Furthermore, one may consider extracting labels from incorrectly decoded blocks, by keeping specific uncoded symbols for which one has a high level of confidence. We leave the study of these extensions for future work.

 	
	\vspace{-0.2cm}
	\section{Numerical Evaluations}
\label{sec:Simulation}
\vspace{-0.1cm} 
We next detail the simulation study used for evaluating Meta-ViterbiNet. The source code used in our experiments is available at https://github.com/tomerraviv95/MetaViterbiNet. 

\vspace{-0.4cm}
\subsection{Evaluated Equalizers}
\label{subsec:simEqualizers}
\vspace{-0.1cm}
In order to evaluate Meta-ViterbiNet, we have implemented the following detectors. 

\subsubsection{Equalizers}
We consider two  \ac{dnn}-aided receivers:
\begin{itemize}
    \item The ViterbiNet equalizer detailed in Subsection~\ref{subsec:ViterbiNet}, whose internal  \ac{dnn} is implemented using three fully-connected layers of sizes $1\times 100$, $100\times 50$, and $50 \times \CnstSize^\Mem$, with activation functions set to sigmoid (after first layer), ReLU (after second layer), and softmax output layer.
    \item A recurrent neural network  symbol detector, comprised of a sliding-window \ac{lstm} classifier with two hidden layers of 256 cells and window size $\Mem$, representing a black-box \ac{dnn} benchmark \cite{tandler2019recurrent}.
\end{itemize}

\subsubsection{Training Methods}
Before the evaluation phase begins, we generate a set $\mathcal{D}_0$ of $T_t$ pilot blocks. We then use the following methods for adapting the deep equalizers:
\begin{itemize}
    \item {\em Joint training}: The \ac{dnn} is  trained on $\mathcal{D}_0$ only by minimizing the empirical cross-entropy loss, and no additional training is done in the evaluation phase.
    \item {\em Online training} \cite{shlezinger2019viterbinet}: The \ac{dnn} is initially trained  on $\mathcal{D}_0$ by minimizing the empirical cross-entropy loss. Then, during evaluation, the \ac{dnn} parameters vector $\Weights_j$ is re-trained on each successfully decoded data block and on each incoming pilot block. Precisely, online training follows \eqref{eq:local_update} by using $\Weights_{j}$ in lieu of $\HyperParams_{j+1}$. 
    \item {\em Online meta-learning}: 
    Here, we first meta-train $\HyperParams_0$ with $\mathcal{D}_0$ similar to \eqref{eq:meta_2} as
    \begin{align*}
     \HyperParams_{0} = \mathop{\arg \min}_{\HyperParams}\!\!\! \sum_{\{\myVec{s}^\Blklen_{\hat{j}+1}, {\myVec{y}}_{\hat{j}+1}^\Blklen\} \in \mathcal{D}_0}\!\!\! \mySet{L}_{\hat{j}+1}(\Weights_{\hat{j}+1}\!= \HyperParams \!-\! \kappa \nabla_{\HyperParams}\mySet{L}_{\hat{j}}(\HyperParams)).
\end{align*}
    
	This  process yields the initial hyperparameters $\HyperParams_0$. 
	Then, during evaluation,   Algorithm~\ref{alg:online-meta-learning} is used with online learning every block and online meta-learning every $K=5$ blocks. The number of online meta-learning updates   equals that of online training, thus inducing a relative small overhead due to its additional computations.
\end{itemize}

\vspace{-0.75cm}
\subsection{Simulation Results}
\label{subsec:simulation_results}
\vspace{-0.1cm}
The combination of ViterbiNet equalizer and online meta-training corresponds to the proposed Meta-ViterbiNet. Recalling Figure \ref{fig:transmission}, frames consist of $t_f = 25$ blocks, i.e., each pilots block is followed by $24$ coded data blocks. The messages   are encoded using a \acl{rs} [17,15] code with two parity symbols. Thus, each message $\myVec{m}$ is comprised of $(\Blklen - 16)$ bits under \acl{bpsk} modulation, i.e., $\mySet{S} = \{\pm 1\}$. 

We consider a linear Gaussian channel, whose input-output relationship is given by
\begin{equation}
\label{eqn:Gaussian}
\myY_{i,j} = \sum_{l=0}^{\Mem-1} h_{l,j}S_{i-l,j} + w_{i,j},
\end{equation}
where $\myVec{h}_j = [h_{0,j},\ldots,h_{\Mem-1,j}]^T$ are the real channel taps, and $w_{i,j}$ is \acl{awgn} with variance $\sigma^2$. We set channel memory to $\Mem = 4$ with the taps $\{h_{l,j}\}$ being generated using a synthetic model representing oscillations of varying frequencies, as well as using the COST 2100 model for indoor wireless communications \cite{liu2012cost}. 

\subsubsection{Synthetic Channel}
In the first experiment we consider a synthetic periodically time-varying channel. Here,  the signals received during the pilots used for initial training $(\mathcal{D}_0)$ are subject to the time-varying channel whose taps are illustrated in Fig.~\ref{fig:synthetic_train_channel}; while we use the taps illustrated in Fig.~\ref{fig:synthetic_test_channel} for the rest of the experiment. This channel presents oscillations of varying frequencies, where  the periods of the taps become aligned as the noise subsides. We set the block length to $\Blklen = 136$ symbols, representing a relatively short coherence duration for the time-varying channel. 

\begin{figure*}
    \centering
    \begin{subfigure}[b]{0.48\textwidth}
    \includegraphics[width=\textwidth,height=0.22\textheight]{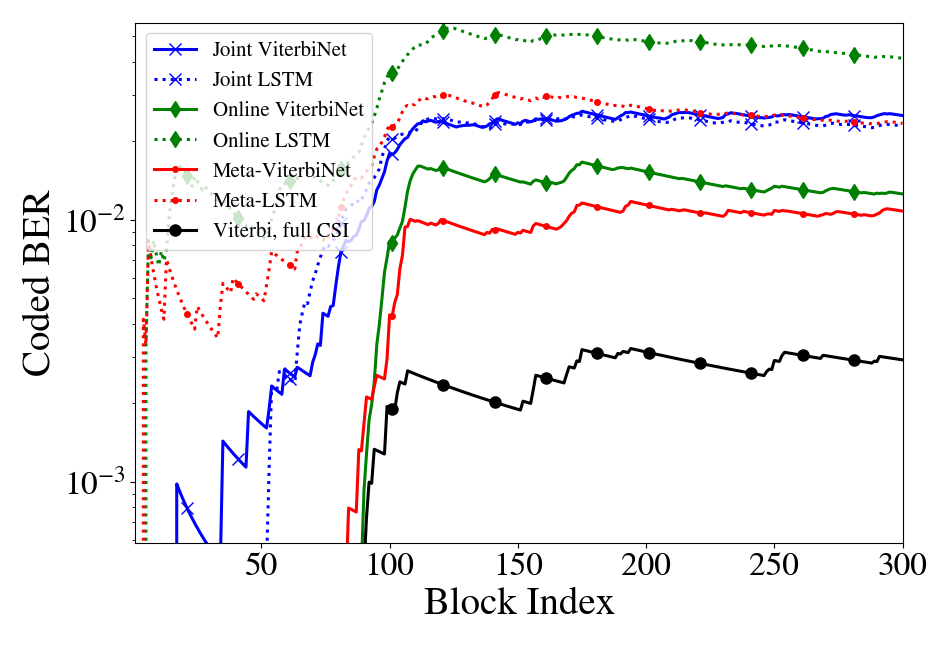}
    \caption{Coded \ac{ber} vs. block index, ${\rm SNR} = 12$ dB.}
    \label{fig:SyntheticBERvsBlock}
    \end{subfigure}
    \begin{subfigure}[b]{0.48\textwidth}
    \includegraphics[width=\textwidth,height=0.22\textheight]{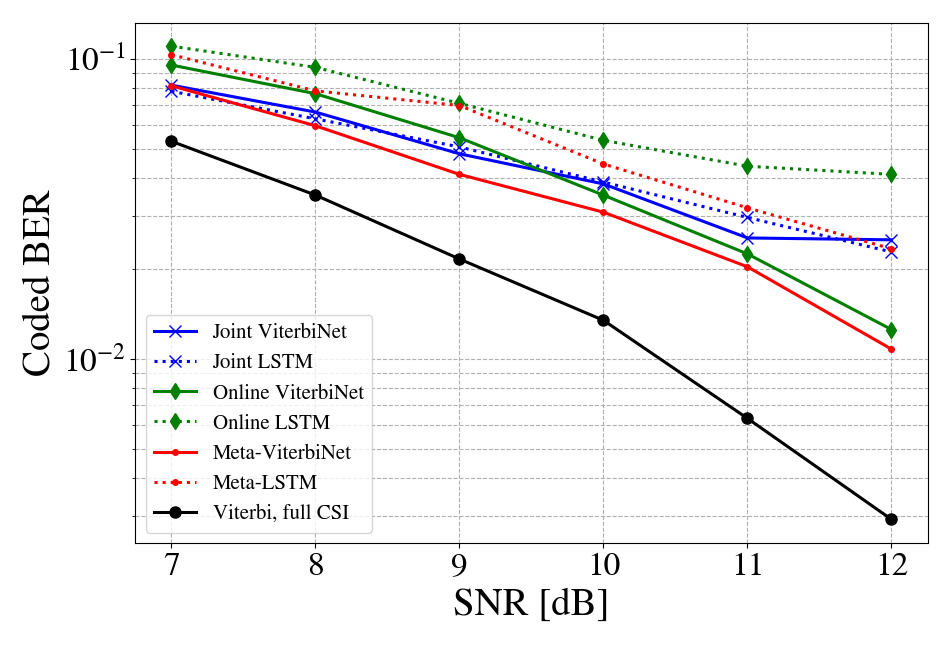}
    \caption{Coded \ac{ber} after 300 blocks vs. SNR.}
    \label{fig:SyntheticBERvsSNR}
    \end{subfigure}
    \caption{Synthetic linear Gaussian channel, $\Blklen = 136$.}
    \label{fig:SyntheticBER} 
    \figSpace
\end{figure*}

In Fig.~\ref{fig:SyntheticBERvsBlock} we plot the evolution of the average coded \ac{ber} of the considered receivers when the \ac{snr}, defined as $1/\sigma^2$, is set to $12$ dB. Fig.~\ref{fig:SyntheticBERvsBlock} shows that Meta-ViterbiNet significantly outperforms its benchmarks. In particular, it is demonstrated that each of the ingredients combined in Meta-VitebiNet facilitates operation in time-varying conditions: The ViterbiNet architecture consistently outperforms the black-box \ac{lstm} classifier; Online training yields reduced \ac{ber} as compared to joint learning; and its combination with meta-learning yields the lowest \ac{ber}.  

To further validate that these gains also hold for different \acp{snr}, we show in Fig.~\ref{fig:SyntheticBERvsSNR} the average coded \ac{ber} of the evaluated receivers after 300 blocks. We observe in Fig.~\ref{fig:SyntheticBERvsSNR} that for \ac{snr} values larger than $8$ dB, Meta-ViterbiNet consistently achieves the lowest  \ac{ber} values among all considered data-driven receivers, with gains of up to 0.5dB.

\begin{figure*}
    \centering
    \begin{subfigure}[b]{0.48\textwidth}
    \includegraphics[width=\textwidth,height=0.22\textheight]{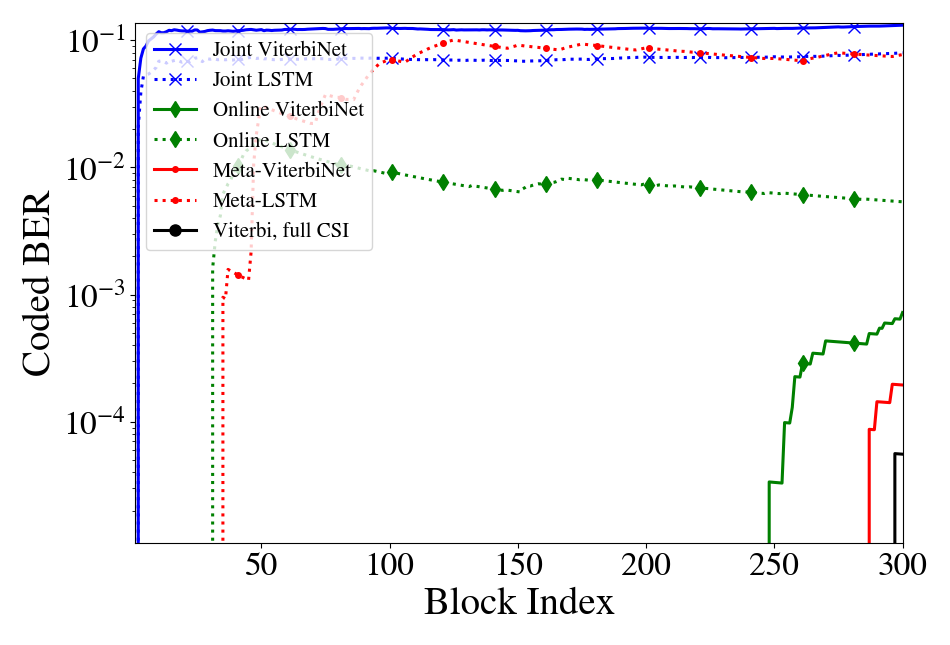}
    \caption{Coded \ac{ber} vs. block index, ${\rm SNR} = 12$ dB.}
    \label{fig:costBERvsBlock}
    \end{subfigure}
    \begin{subfigure}[b]{0.48\textwidth}
    \includegraphics[width=\textwidth,height=0.22\textheight]{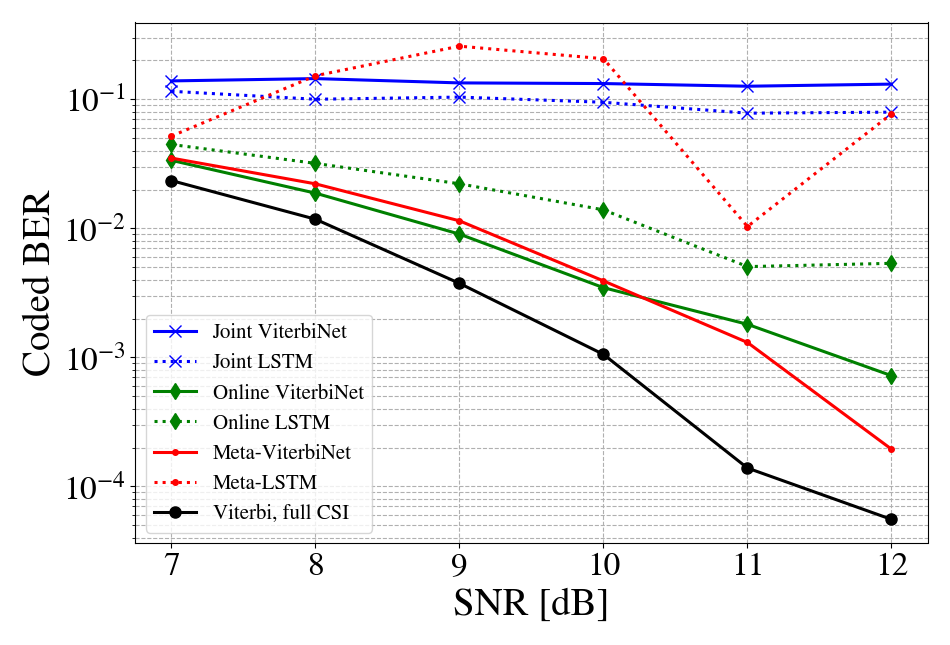}
    \caption{Coded \ac{ber} after 300 blocks vs. SNR.}
    \label{fig:costBERvsSNR}
    \end{subfigure}
    \caption{COST 2100 channel, $\Blklen = 136$.}
    \label{fig:CostBER} 
    \figSpace
\end{figure*}

\subsubsection{COST 2100 Channel}
Next, we consider channels generated using the COST 2100 geometry-based stochastic channel model \cite{liu2012cost}. In particular, we generate each realization of the taps using an indoor hall $5$GHz setting with single-antenna elements. We use the same block length and number of error-correction symbols, as well as the same initial training set $\mySet{D}_0$ as in the synthetic model. The test is carried out using a sequence of difference realizations illustrated in Fig.~\ref{fig:COSTChannel}. This setting may represent a user moving in an indoor setup while switching between different microcells. Succeeding on this scenario requires high adaptivity since there is considerable mismatch between the train and test channels. 

In Fig.~\ref{fig:costBERvsBlock} we illustrate the time evolution of the coded \ac{ber} of the compared receivers for \ac{snr} of $12$ dB. Fig.~\ref{fig:costBERvsBlock} demonstrates the ability of Meta-ViterbiNet to operate reliably in time-varying channel conditions, while improving upon both ViterbiNet without meta-learning, as well as over conventional data-driven architectures based on \ac{lstm}. Fig.~\ref{fig:costBERvsSNR} plots the average coded \ac{ber} after $300$ blocks versus \ac{snr}, showing that Meta-ViterbiNet achieves an improvement of up to 0.6dB. 

	\vspace{-0.4cm}
	\section{Conclusions}
	\label{sec:Conclusions}
	\vspace{-0.1cm}
We proposed Meta-ViterbiNet, a data-driven symbol detector with meta-learned hyperparameter vector designed to track channel variations via online training. Meta-ViterbiNet incorporates three key ingredients that enable the tracking of rapidly time-varying channels: a model-based \ac{dnn} architecture; an online adaptation scheme with optimized inital weights; and the use of coded data blocks for self-supervised training. Numerical study demonstrates that, by properly integrating these methods, Meta-ViterbiNet is capable of outperforming previous \ac{dnn}-aided receivers.  
	\vspace{-0.5cm}
	\bibliographystyle{IEEEtran}
	\bibliography{IEEEabrv,refs}
\end{document}